\documentclass[conference]{IEEEtran}
\IEEEoverridecommandlockouts
\usepackage{cite}
\usepackage{amsmath,amssymb,amsfonts}
\usepackage{algorithmic}
\usepackage{graphicx}
\usepackage{textcomp}
\usepackage{xcolor}
\usepackage{caption, subcaption}

\usepackage[ruled,linesnumbered,noend]{algorithm2e}
\usepackage{multirow}
\usepackage{threeparttable}
\usepackage{makecell}
\usepackage{pifont}
\usepackage{bm}
\usepackage{amsmath}
\usepackage{amsthm}
\usepackage{hyperref}
\usepackage{url}

\usepackage{booktabs}

\def\BibTeX{{\rm B\kern-.05em{\sc i\kern-.025em b}\kern-.08em
    T\kern-.1667em\lower.7ex\hbox{E}\kern-.125emX}}
\begin{document}

\title{ 
Fast Authenticated and Interoperable Multimedia Healthcare Data over Hybrid-Storage Blockchains\\
\thanks{*Corresponding author: Haiqin Wu. This work was supported in part by the National Key R\&D
 Program of China under Grant 2022YFB2701400, and in part by the National Natural Science Foundation of China under Grants 62202167, 62172162, and 62132005.}
}


\author{
\IEEEauthorblockN{
Jucai Yang,
		Liang Li, 
		Yiwei Gu,
        and Haiqin Wu\IEEEauthorrefmark{1}}
\IEEEauthorblockA{East China Normal University, Shanghai, China }
\IEEEauthorblockA{Email: \{jcyang, liangli, ywgu\}@stu.ecnu.edu.cn, hqwu@sei.ecnu.edu.cn}

}
\maketitle

\begin{abstract}
The integration of blockchain technology into healthcare presents a paradigm shift for secure data management, enabling decentralized and tamper-proof storage and sharing of sensitive Electronic Health Records (EHRs). 
However, existing blockchain-based healthcare systems, while providing robust access control, commonly overlook the high latency in user-side re-computation of hashes for integrity verification of large multimedia data, impairing their practicality, especially in time-sensitive clinical scenarios. 
In this paper, we propose FAITH, an innovative scheme for \underline{F}ast \underline{A}uthenticated and \underline{I}nteroperable\footnote{We focus on the secure data sharing across healthcare institutions.} mul\underline{T}imedia \underline{H}ealthcare data storage and sharing over hybrid-storage blockchains. Rather than user-side hash re-computations, FAITH lets an off-chain storage provider generate verifiable proofs using recursive Zero-Knowledge Proofs (ZKPs), while the user only needs to perform lightweight verification. For flexible access authorization, we leverage Proxy Re-Encryption (PRE) and enable the provider to conduct ciphertext re-encryption, in which the re-encryption correctness can be verified via ZKPs against the malicious provider. All metadata and proofs are recorded on-chain for public verification. 
We provide a comprehensive analysis of FAITH's security regarding data privacy and integrity. We implemented a prototype of FAITH, and extensive experiments demonstrated its practicality for time-critical healthcare applications, dramatically reducing user-side verification latency by up to $98\%$, bringing it from $4$ s down to around $70$ ms for a $5$ GB encrypted file. 
\end{abstract}


\section{Introduction}
The application of blockchain in healthcare is reshaping conventional isolated systems, with hybrid-storage architectures~\cite{lv2024cross, wu2023patientcentered, HASAN2025110462} emerging as a prevailing strategy to scale medical services. In this model, large raw electronic health records (EHRs) are stored in an off-chain service provider (SP), while only the metadata (e.g., hashes) are recorded on-chain. However, this model introduces significant security concerns as the off-chain SP is often untrusted, posing risks to data privacy and integrity \cite{episourceBreach2025}. In data sharing scenarios, clinicians or researchers as data users request patients’ (i.e., data owners’) EHRs for further treatment or academic studies, and should be able to check the integrity of data received from the SP.

 For secure data storage and sharing in healthcare scenarios, existing blockchain-based solutions commonly employ cryptographic primitives like Proxy Re-Encryption (PRE)~\cite{lv2024cross, raghav2023privacypreserving, feng2023privacy,HASAN2025110462} or Attribute-Based Encryption (ABE)~\cite{han2025hybrid, xu2024privacypreserving, wu2023patientcentered, tao2024care, fang2020Privacy} for flexible access delegation or fine-grained access control over encrypted EHRs, respectively. However, they work in an honest-but-curious model and overlook the SP's misbehavior (e.g., incorrect re-encryption). Some studies~\cite{lv2024cross, wu2023patientcentered, HASAN2025110462} adopted a hash re-computation-based approach for data integrity checking, which unfortunately requires users to hash entire large multimedia files (e.g., MRIs) to verify their integrity. The incurred significant latency is unsuitable for time-sensitive scenarios such as emergency care where delays can have severe consequences.

To address the above issues, we introduce FAITH, a novel system for \underline{F}ast \underline{A}uthenticated and \underline{I}nteroperable mul\underline{T}imedia \underline{H}ealthcare data storage and sharing over hybrid-storage blockchains. FAITH overcomes the performance bottleneck of integrity verification for large EHRs. Our core innovation is to transform user-side intensive hash re-computation of large data to a lightweight proof verification for data integrity checking, in which the proof generation is offloaded to the SP who proves the integrity of its stored data based on recursive Zero-Knowledge Proofs (ZKPs). We further incorporate PRE for flexible access authorization and data sharing, and anyone can verify the off-chain ciphertext re-encryption with ZKPs recorded on-chain. 
Overall, FAITH provides a holistic solution for efficiently verifiable and interoperable secure sharing of large multimedia EHRs among patients and healthcare institutions. This caters to time-critical healthcare scenarios. Our main contributions are summarized below:
\begin{enumerate}
    \item We are the first to identify and resolve the key performance bottleneck of user-side integrity verification for large multimedia files in blockchain healthcare systems.
    \item We propose FAITH, an interoperable healthcare architecture that enables fast integrity verification by shifting the burden from user-side hash re-computation to lightweight verification of on-chain ZKPs.
    \item We demonstrate FAITH's security guarantees and performance, achieving around a 98\% reduction in verification time for a $5$~GB file compared to hash recomputation.
\end{enumerate}

\section{Preliminaries}
\label{Sec:Pre} 

\subsection{Proxy Re-Encryption}
\label{Sec:PRE}
Proxy Re-Encryption (PRE) is an advanced public-key encryption that enables a proxy to transform a ciphertext from one key to another without learning the underlying plaintext, making it ideal for secure data sharing~\cite{ateniese2006improved}. A PRE scheme consists of the following algorithms:
\begin{itemize}
    \item \textsf{PRE.KeyGen$(1^\lambda)$}$\rightarrow(pk, sk)$: Generates a key pair.
    
    \item \textsf{PRE.Enc($pk_A, m$)}$\rightarrow c_A$: Encrypts a message $m$ under public key $pk_A$ to get ciphertext $c_A$.
    
    \item \textsf{PRE.ReKeyGen($sk_A, pk_B$)}$\rightarrow rk_{A \to B}$: Generates a re-encryption key $rk_{A \to B}$ using $sk_A$ and $pk_B$.
    
    \item \textsf{PRE.ReEnc($rk_{A \to B}, c_A$)}$\rightarrow c_B$: Uses $rk_{A \to B}$ to transform ciphertext $c_A$ into a new ciphertext $c_B$.
    
    \item \textsf{PRE.Dec($sk_B, c_B$)}$\rightarrow m$: Decrypts $c_B$ with secret key $sk_B$ to recover message $m$.
\end{itemize}

\subsection{Recursive Zero-Knowledge Proofs}
\label{Sec:ZKP}
A Zero-Knowledge Proof (ZKP)~\cite{goldwasser2019knowledge} allows a prover to prove a statement's validity without revealing secret information. A Zero-Knowledge Succinct Non-interactive Argument of Knowledge (zk-SNARK) is a highly efficient ZKP that generates a succinct proof. A zk-SNARK scheme includes:
\begin{itemize}
    \item \textsf{ZKP.Setup($\mathcal{C}$)}$\rightarrow (prk,vrk)$: Generates a proving key $prk$ and a verification key $vrk$ for a given circuit $\mathcal{C}$.
    
    \item \textsf{ZKP.Prove($prk, x, w$)}$\rightarrow \pi$: Using a private witness $w$, generates a proof $\pi$ for a public input $x$.
    
    \item \textsf{ZKP.Verify($vrk, x, \pi$)}$\rightarrow 0/1$: Verifies the proof $\pi$.
\end{itemize}
Recursive ZKPs~\cite{PolygonZero2022Plonky2} extend this by allowing one ZKP to verify other ZKPs. Since a zk-SNARK verifier itself is a circuit, its execution can be proven within another zk-SNARK. This enables aggregating multiple proofs $\{\pi_1, \dots, \pi_n\}$ into a single, compact proof that attests to their collective validity.

\section{Construction}
\label{sec:construction} 

\subsection{System Model}
\label{sec:systemModel}
As illustrated in Fig.~\ref{fig_systemModel}, our proposed system, FAITH, consists of five main entities:

\begin{figure}[htbp]
\centerline{\includegraphics[width=0.5\textwidth]{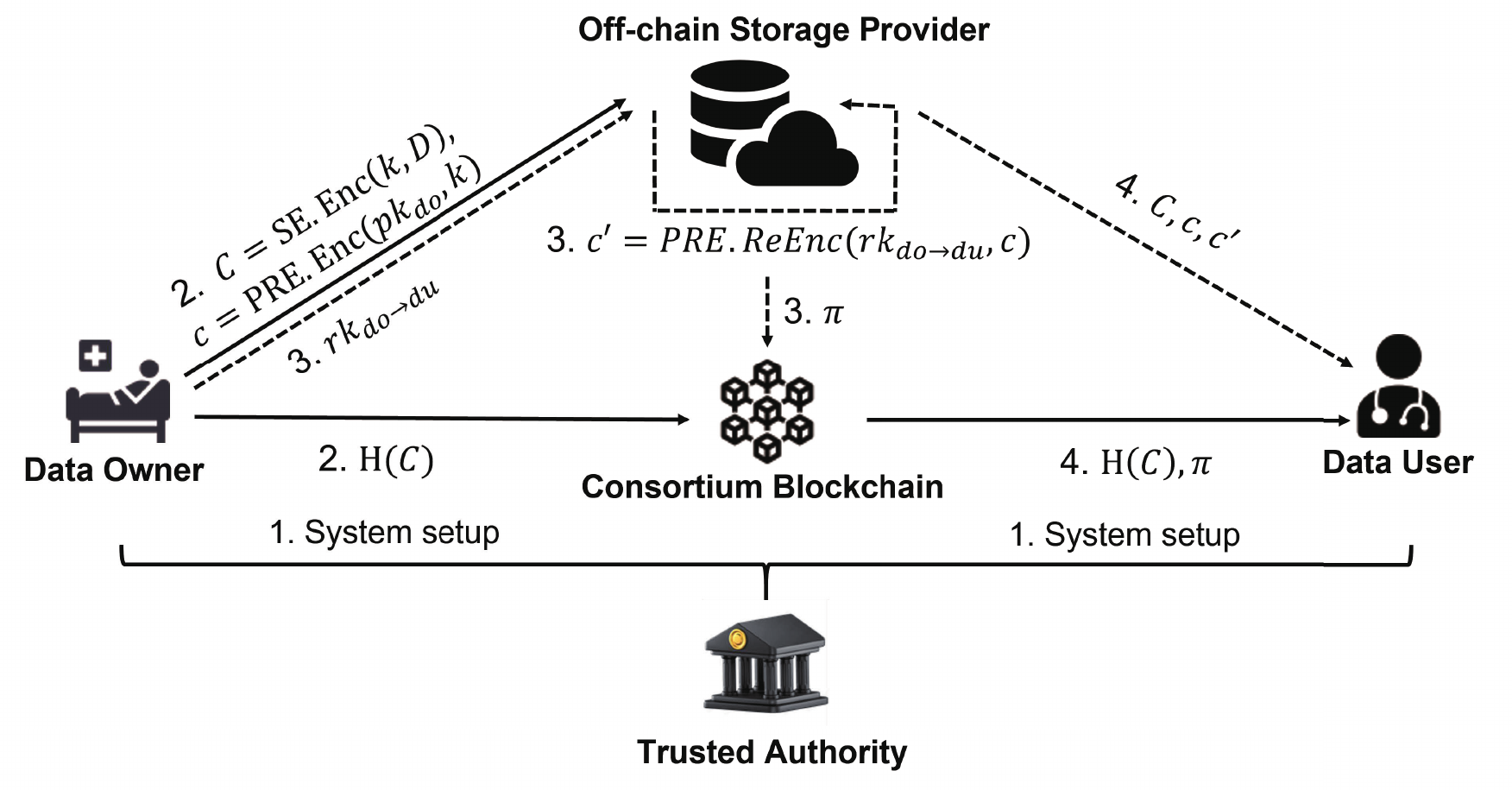}}
\caption{System model of FAITH.}
\label{fig_systemModel}
\end{figure}

\begin{itemize}
    \item \textbf{Trusted Authority (TA):} The TA is a trusted third party (government agency or healthcare authority) responsible for the system setup and public parameter generation.
    
    \item \textbf{Data Owner (DO):} A patient who encrypts their multimedia healthcare data and grants access rights to others.

    \item \textbf{Data User (DU):} A clinician or researcher who requests authorized access to the DO's data.

    \item \textbf{Off-chain Storage Provider (SP):} The SP stores large encrypted data, performs proxy re-encryption upon request, and is tasked with generating ZKPs for integrity.

    \item \textbf{Consortium Blockchain:} A tamper-proof ledger for storing metadata (e.g., data hashes) and verifiable proofs.
\end{itemize}

The overall workflow of FAITH is as follows. First, the TA conducts system setup, including public parameter generation and cryptographic primitive definition (step 1). Then the DO encrypts a large multimedia EHR $D$ with a symmetric key $k$, and derives a ciphertext $C$. The symmetric key $k$ itself is encrypted using a PRE scheme under the DO's public key $pk_{o}$, which generates an encrypted key $c$. Both $C$ and $c$ are sent to the SP while the metadata $H(C)$ is recorded on-chain (step 2).  
When a DU requests access to the EHR, the DO generates a re-encryption key $rk_{o \to u}$ and sends it to the SP, who then transforms $c$ into a new ciphertext $c'$ with $rk_{o \to u}$. Meanwhile, it generates a zero-knowledge proof $\pi$ certifying the honest execution of re-encryption and data storage integrity, 
$\pi$ is anchored to the blockchain (step 3). Then SP returns the corresponding result to the DU. With the on-chain recorded $H(C)$ and proof $\pi$, the DU finally performs an integrity check and then recovers the original EHR $D$ (step 4).

\subsection{Scheme Implementation}
The FAITH system is constructed through four main phases: setup, data upload, access granting, and data retrieval.

\textbf{1. System Setup.}
The TA performs a one-time setup, defining system-wide parameters and cryptographic primitives: a security parameter $\lambda$, bilinear map parameters $\{\mathbb{G}_1, \mathbb{G}_2, p, e, g_1, g_2\}$, a hash function $\textsf{H}: \{0,1\}^* \to \mathbb{Z}^*_p$, a symmetric encryption scheme $\textsf{SE} = \{\textsf{SE.KeyGen}(1^\lambda), \textsf{SE.Enc}(k,D), \textsf{SE.Dec}(k,C)\}$, a proxy re-encryption scheme $\textsf{PRE}$ (see Section~\ref{Sec:PRE}), and a zero-knowledge proof system $\textsf{ZKP}$ (see Section~\ref{Sec:ZKP}). The TA generates proving and verification keys for three ZKP circuits: $\{prk_\text{int}, vrk_\text{int}\} \leftarrow \textsf{ZKP.Setup}(\mathcal{C}_\text{int}, 1^\lambda)$ for data integrity, $\{prk_\text{pre}, vrk_\text{pre}\} \leftarrow \textsf{ZKP.Setup}(\mathcal{C}_\text{pre}, 1^\lambda)$ for re-encryption correctness, and $\{prk_\text{agg}, vrk_\text{agg}\} $ $ \leftarrow \textsf{ZKP.Setup}(\mathcal{C}_\text{agg}, 1^\lambda)$ for proof aggregation. The details of \textsf{ZKP.Setup} construction can be found in \cite{PolygonZero2022Plonky2}. Each user (DO or DU) generates their own PRE key pair by running $\textsf{PRE.KeyGen}(1^\lambda)$. Specifically, s/he chooses two random numbers $sk_1,sk_2 \in \mathbb{Z}^*_p$ as secret key $sk = (sk_1, sk_2)$ and the public key is generated as $pk = (g_2^{sk_1}, g_1^{sk_2})$. To improve clarity, we denote the DO's key pair as $(pk_o, sk_o)$ where $sk_o=(o_1, o_2)$ and $pk_o=(g_2^{o_1}, g_1^{o_2})$, and the DU's key pair as $(pk_u, sk_u)$ where $sk_u=(u_1, u_2)$ and $pk_u=(g_2^{u_1}, g_1^{u_2})$.

\textbf{2. Data Encryption and Upload.}
To securely store a multimedia file $D$, the DO performs the following:
\begin{enumerate}
    \item Encrypts the file $D$ with a fresh symmetric key $k \leftarrow \textsf{SE.KeyGen}(1^\lambda)$ to get ciphertext $C \leftarrow \textsf{SE.Enc}(k, D)$.
    \item Encrypts the key $k$ using their own public key $pk_o$, resulting in a ciphertext $c \leftarrow \textsf{PRE.Enc}(pk_o, k)$, where
    \begin{equation}
        c = (c_1, c_2), c_1 = g_1^r, \; c_2  =k \cdot (g_2^{o_1})^r = k \cdot g_2^{o_1 r}.
    \end{equation}
    \item Records the hash of the data ciphertext, $h \leftarrow \textsf{H}(C)$, on the blockchain and sends the tuple $\langle id, C, c \rangle$ to SP.
\end{enumerate}

Upon receipt, the SP generates an initial integrity proof $\pi_\text{int} \leftarrow \textsf{ZKP.Prove}(prk_\text{int}, x_\text{int}, w_\text{int})$ which attests that it stores the correct data corresponding to the on-chain hash $h$, where $x_\text{int} \leftarrow h$ is a public statement, and $w_\text{int} \leftarrow C$ is the witness. Then the SP stores $\langle id, C, c, \pi_\text{int} \rangle$ in its storage.

\textbf{3. Access Granting and Proof Aggregation.}
When a DO grants access to a DU, the following steps occur:
\begin{enumerate}
    \item The DO generates a re-encryption key $rk_{o \to u} \leftarrow \textsf{PRE.ReKeyGen}(sk_o, pk_u)$ where $rk_{o \to u} = (g_1^{u_2})^{o_1} = g_1^{o_1 u_2}$, and sends it to the SP.
    \item The SP uses $rk_{o \to u}$ to transform the key ciphertext $c$ into a re-encrypted version $c'$ for the DU, where
    \begin{align}
        c_1' &= e(c_1, rk_{o \to u}) =e(g_1^r, g_1^{o_1 u_2}) = g_2^{r o_1 u_2}, \nonumber \\
        c_2'& = c_2 = k \cdot g_2^{o_1 r}.
    \end{align}
    \item The SP generates a proof $\pi_\text{pre}$ certifying the correctness of the re-encryption operation:
    \begin{equation}
        \pi_\text{pre} \leftarrow \textsf{ZKP.Prove}(prk_\text{pre}, x_\text{pre}, w_\text{pre}),
    \end{equation}
    where  $x_\text{pre} \!\leftarrow\! c'\|c$ is public and $w_\text{pre} \leftarrow rk_{{o} \to {u}}$ is private.
    \item Critically, the SP uses a recursive ZKP to aggregate the integrity proof $\pi_\text{int}$ and the re-encryption proof $\pi_\text{pre}$ into a single, compact proof $\pi_\text{agg}$:
    \begin{equation}
    \label{eq:agg}
        \pi_\text{agg} \leftarrow \textsf{ZKP.Prove}(prk_\text{agg}, x_\text{agg}, w_\text{agg}),
   \end{equation}
    where where $x_\text{agg} \leftarrow h \| c' \| c$ and $w_\text{agg} \leftarrow C \| rk_{o\to u}$. This aggregated proof is then published on the blockchain.
\end{enumerate}

\textbf{4. Data Request and Decryption.}
To access the data, the DU sends a query $Q = \langle id \rangle$ to SP who then returns the tuple $\langle id,  C, c, c', \pi_\text{agg} \rangle$. The DU then performs the following steps:
\begin{enumerate}
    \item Fetches the data hash $h$ and the aggregated proof $\pi_\text{agg}$ from the blockchain, and performs a single, lightweight verification \textsf{ZKP.Verify}$(vrk_\text{agg}, x_\text{agg}, \pi_\text{agg})$. This check simultaneously validates both the integrity of the file $C$ and the correctness of the key re-encryption.
    \item If verification succeeds, the DU uses their secret key $sk_u$ to decrypt $c'$ to recover the symmetric key $k$:
    \begin{align}
    \label{eq:decrypt}
        c'_2 \cdot (c'_1)^{-1/u_2} 
        &= (k \cdot g_2^{o_1 r}) \cdot (g_2^{r o_1 u_2})^{-1/u_2} \nonumber \\
        &= (k \cdot g_2^{o_1 r}) \cdot (g_2^{-r o_1})=k,
    \end{align}
    and then decrypts $C$ to obtain $D \leftarrow \textsf{SE.Dec}(k,C)$.
\end{enumerate}

This process can be extended to handle multiple files, where proofs for all files can be combined into a single aggregated proof, further reducing the verification overhead for the DU.

 \begin{table}[h]
\centering
\caption{Performance of SE encryption and decryption.}
\label{tab:aes-performance-simple}
\renewcommand{\arraystretch}{1.2} 
\begin{tabular}{|l|c|c|c|c|c|}
\hline
\textbf{Message size (GB)}    & 1       & 2       & 3       & 4        & 5        \\ \hline
\textbf{SE.Enc (s)} & 2.800   & 5.604   & 8.381   & 11.167   & 13.967   \\ \hline
\textbf{SE.Dec (s)} & 0.820   & 1.656   & 2.482   & 3.249    & 4.0527    \\ \hline
\end{tabular}
\end{table}

\begin{table}[h]
\centering
\caption{Time cost of PRE algorithms.}
\label{tab:pre-performance-simple}
\begin{tabular}{@{}lcccccc@{}}
\toprule
\textbf{PRE algorithms}  & \textbf{KeyGen} & \textbf{ReKeyGen} & \textbf{Enc } & \textbf{ReEnc} & \textbf{Dec} \\ 
\midrule
\textbf{Time cost (ms)}  & 3.135 & 0.292 & 1.506 & 2.615 & 3.791 \\ 
\bottomrule
\end{tabular}
\end{table}

\begin{figure*}[t]
    \centering
    \begin{subfigure}[t]{0.3\textwidth}
        \centering
        \includegraphics[width=1.00\textwidth]{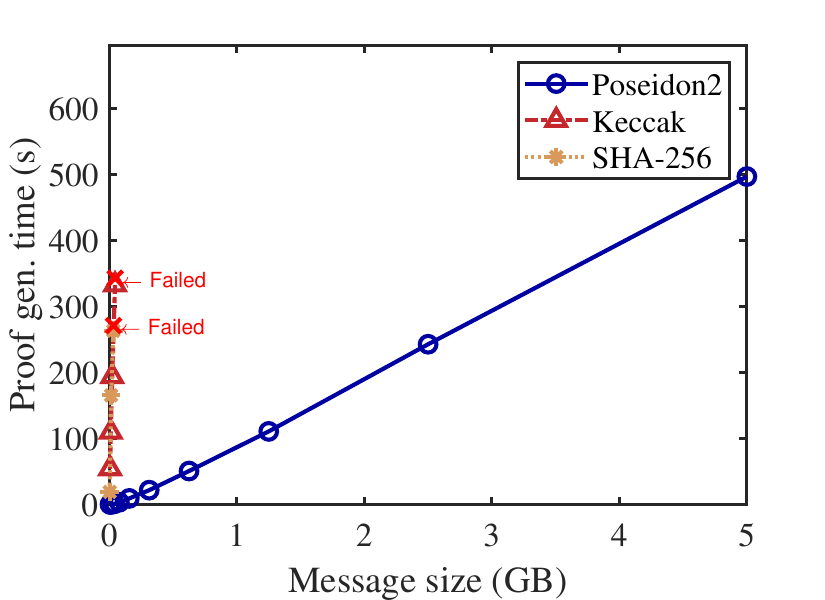}
        \caption{Proof generation time}
        \label{fig: Proof generation time}
    \end{subfigure}
    \hfill
    \begin{subfigure}[t]{0.3\textwidth}
        \centering
        \includegraphics[width=1.00\textwidth]{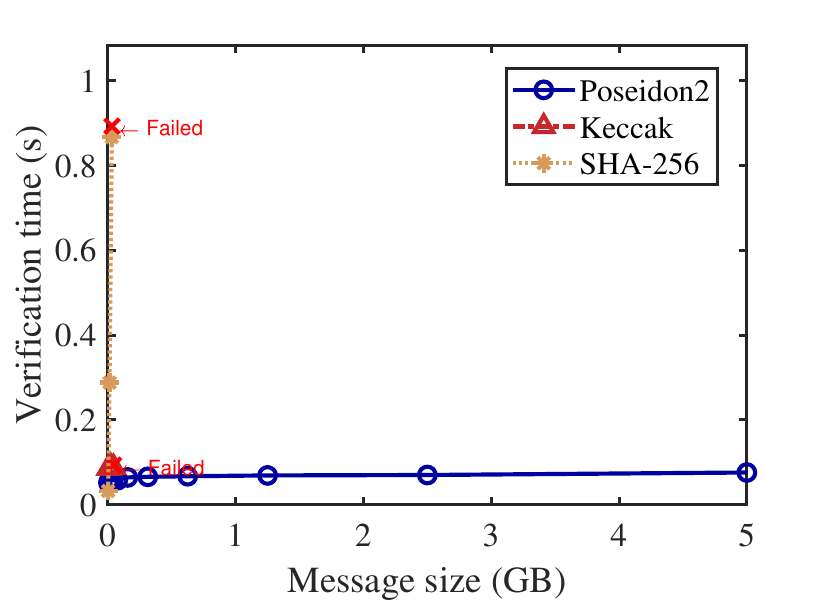}
        \caption{Verification time}
        \label{fig: Verification time}
    \end{subfigure}
    \hfill
    \begin{subfigure}[t]{0.288\textwidth}
        \centering
        \includegraphics[width=1.00\textwidth]{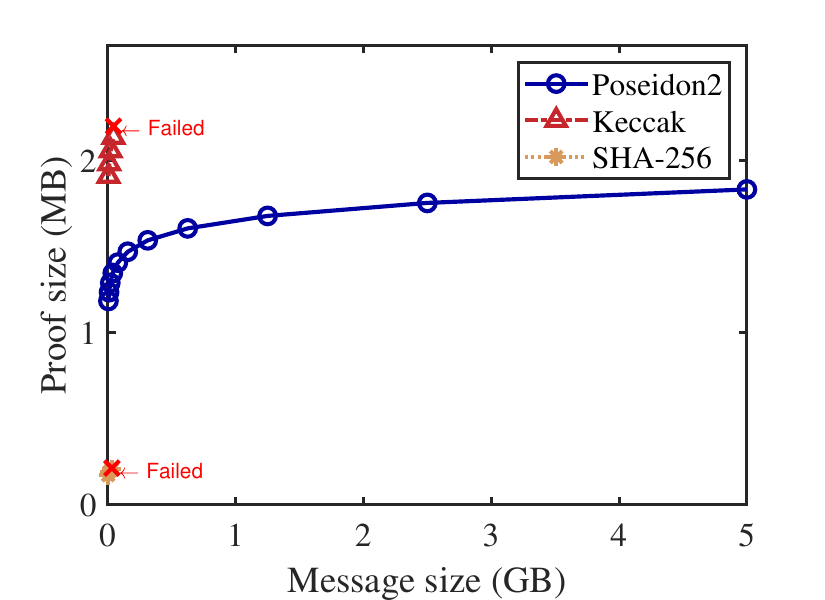}
        \caption{Proof size}
        \label{fig: Proof size}
    \end{subfigure}
    \caption{Comparison of different hash functions under ZKP.}
    \begin{minipage}{\textwidth}
        \small Notes: \textcolor{red}{$\times \leftarrow \text{Failed}$} indicates the proving process terminated at that message size due to errors such as out-of-memory.
    \end{minipage}
    \label{fig: Comparison under ZKP}
\end{figure*}

\begin{figure}[t]
    \centering
    \begin{subfigure}[t]{0.24\textwidth}
        \centering
        \includegraphics[width=1.00\textwidth]{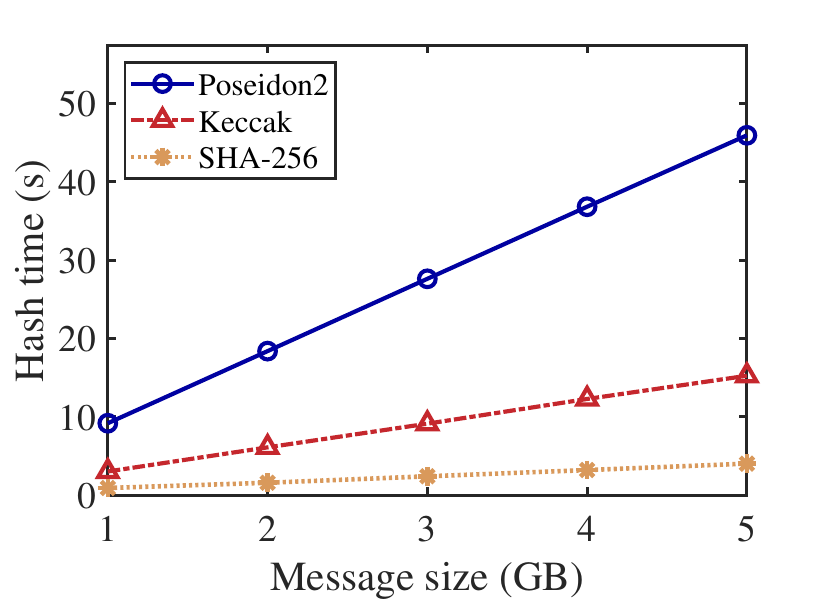}
        \caption{Naive hash re-computation}
        \label{fig: hash time}
    \end{subfigure}
    \hfill
    \begin{subfigure}[t]{0.24\textwidth}
        \centering
        \includegraphics[width=1.00\textwidth]{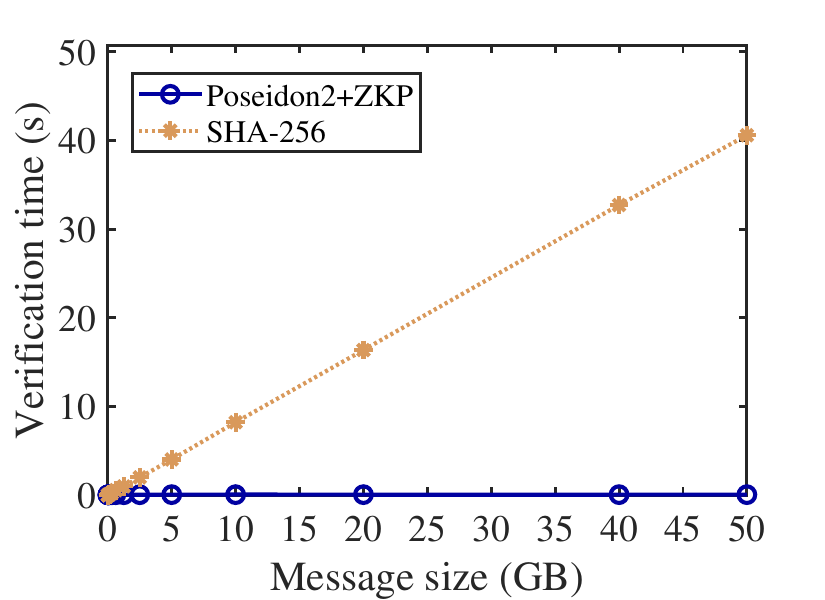}
        \caption{Comparison with benchmark}
        \label{fig: integrity verificatioin}
    \end{subfigure}
    \caption{Integrity verification time under different approaches.}
    \label{fig: comp. of int. ver.}
\end{figure}

\begin{figure}[t]
    \centering
    \begin{subfigure}[t]{0.24\textwidth}
        \centering
        \includegraphics[width=1.00\textwidth]{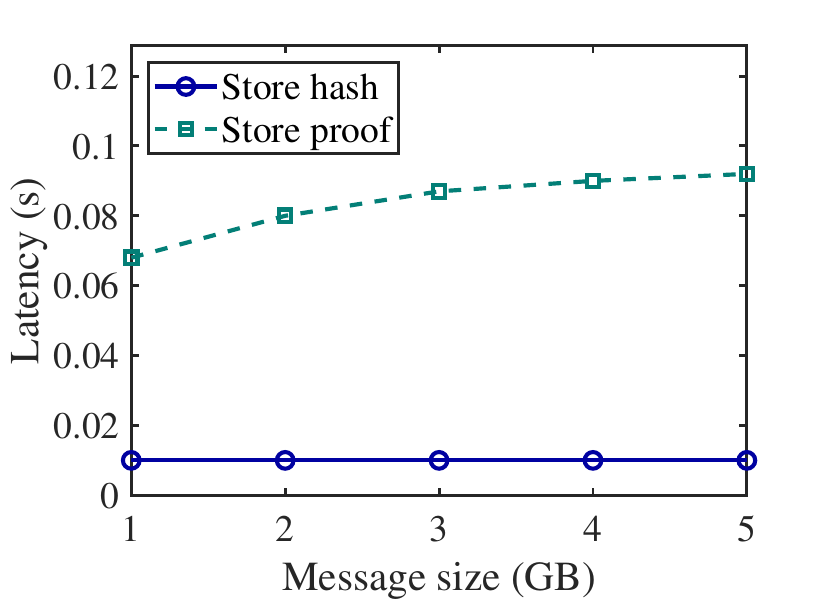}
        \caption{Latency}
        \label{fig: on-chain latency}
    \end{subfigure}
    \hfill
    \begin{subfigure}[t]{0.24\textwidth}
        \centering
        \includegraphics[width=1.00\textwidth]{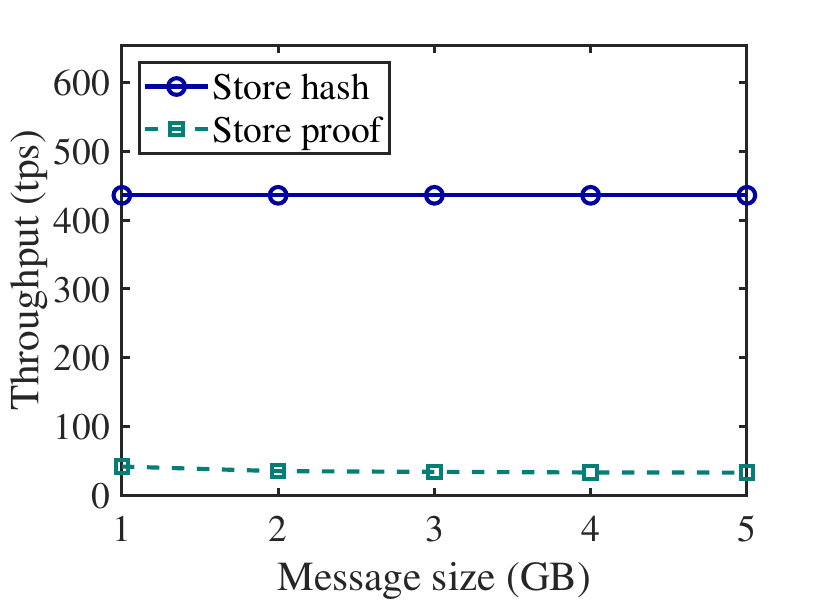}
        \caption{Throughput}
        \label{fig: on-chain throughput}
    \end{subfigure}
    \caption{On-chain performance for storing hash and proof.}
    \label{fig: on-chain performance}
\end{figure}

\section{Evaluation}
\label{sec:evaluation}

\subsection{Security Analysis}
FAITH provides robust security guarantees through its cryptographic design, including data privacy and integrity.

\textbf{Data Privacy.} Privacy is ensured by a two-layer encryption mechanism. First, large multimedia healthcare data is encrypted via a secure symmetric scheme (e.g., AES), rendering it unintelligible to the off-chain SP. Second, the symmetric key itself is encrypted using a PRE scheme. The fundamental security of PRE ensures that only an authorized DU, holding the correct private key, can decrypt the re-encrypted key. The SP, even while performing re-encryption, learns nothing about the key's content. This dual approach provides end-to-end confidentiality, effectively protecting sensitive data from both the SP and any unauthorized parties.

\textbf{Integrity.} FAITH guarantees both data integrity and re-encryption correctness using on-chain hashes and ZKPs. The hash of the large encrypted file is anchored on the immutable blockchain. Instead of requiring the DU to download the entire file for hash re-computation, our system offloads this task to the SP. The SP generates a ZKP attesting that its stored data matches the on-chain hash. Another ZKP is generated to prove the correctness of the PRE operation. Crucially, these proofs are recursively aggregated into a single and compact proof that is stored on-chain. This allows any DU to perform a fast, public, and lightweight verification in milliseconds, regardless of the data size, immediately detecting any malicious tampering or incorrect re-encryption by the SP.

\subsection{Performance Analysis}
We implemented and evaluated a prototype of FAITH using AFGH-06 PRE~\cite{ateniese2006improved}, AES~\cite{daemen1999aes}, and the Plonky2~\cite{PolygonZero2022Plonky2} ZKP system. Our evaluation compared the performance of the ZKP-friendly Poseidon2 hash against traditional hashes like SHA-256 for files up to 5 GB.

The performance of the core cryptographic primitives validates our hybrid design. Symmetric encryption of a 5 GB file takes approximately 14 s, a practical one-time cost, while all PRE operations on the small symmetric key complete in just a few milliseconds (Tables~\ref{tab:aes-performance-simple} \& \ref{tab:pre-performance-simple}).

For the ZKP-based integrity check, our results demonstrate that using a ZKP-friendly hash function like Poseidon2 is critical. As shown in Fig.~\ref{fig: Comparison under ZKP}, our ZKP-based verification time remains near-constant at around 70 ms, regardless of the file size. As observed, Poseidon2 presents the fastest integrity verification performance, merely incurring tens of milliseconds, which is more lightweight than hash recomputation (at least a few seconds). In contrast, the conventional approach of re-computing a SHA-256 hash scales linearly, taking around 4 s for a 5 GB file (see Fig.~\ref{fig: comp. of int. ver.}). Overall, FAITH presents an approximately 98\% verification time reduction by transforming hash re-computation to a near-instantaneous operation. Finally, Fig.~\ref{fig: on-chain performance} depicts the on-chain latency and throughput for our hash and proof storage, confirming FAITH's overall efficiency (the latency was under 0.1 s) and practicality.

\section{Conclusion}
\label{sec:conclusion}


This paper introduced FAITH, a novel framework for fast, interoperable healthcare data sharing that resolves the critical user-side integrity verification bottleneck. It replaces the costly hash re-computation of large files with a lightweight verification of server-generated ZKPs. FAITH also integrates Proxy Re-Encryption for flexible access, using ZKPs to ensure re-encryption correctness against a malicious server. Our analysis confirms robust data privacy and integrity, while experiments demonstrate that FAITH reduces verification time to a constant few milliseconds, irrespective of file size, validating its feasibility for real-world medical applications.

\end{document}